\documentclass{llncs}
\usepackage{makeidx}  
\usepackage{graphicx}
\usepackage{epstopdf}
\usepackage{color}
\usepackage{placeins}
\usepackage{amssymb}
\usepackage{algorithm}
\usepackage[noend]{algpseudocode}
\usepackage{wrapfig}

\newcommand{\tw}{\textrm{tw}}

\begin{document}

\mainmatter

\title{A Solution Merging Heuristic for the Steiner Problem in Graphs Using Tree Decompositions
}

\titlerunning{Solution Merging Heuristic for the STP}        

\author{Thomas Bosman}


\institute{VU University Amsterdam, The Netherlands\\
	\email{t.n.bosman@student.vu.nl}
	}

\date{Received: date / Accepted: date}

\maketitle

\begin{abstract}
Fixed parameter tractable algorithms for bounded treewidth are known to exist for a wide class of graph optimization problems. While most research in this area has been focused on exact algorithms, it is hard to find decompositions of treewidth sufficiently small to make these algorithms fast enough for practical use. Consequently, tree decomposition based algorithms have limited applicability to large scale optimization. However, by first reducing the input graph so that a small width tree decomposition can be found, we can harness the power of tree decomposition based techniques in a heuristic algorithm, usable on graphs of much larger treewidth than would be tractable to solve exactly. We propose a solution merging heuristic to the Steiner Tree Problem that applies this idea. Standard local search heuristics provide a natural way to generate subgraphs with lower treewidth than the original instance, and subsequently we extract an improved solution by solving the instance induced by this subgraph. As such the fixed parameter tractable algorithm becomes an efficient tool for our solution merging heuristic. For a large class of sparse benchmark instances the algorithm is able to find small width tree decompositions on the union of generated solutions. Subsequently it can often improve on the generated solutions fast. 
\end{abstract}

\section{Introduction}
\label{intro}
Treewidth, tree decomposition and related graph decomposition concepts have been studied extensively as a means for finding theoretically efficient algorithms for optimization problems in graphs. For graphs of bounded treewidth, polynomial time algorithms can be found for a large number of graph optimization problems. However, due to large constants hidden in the time complexity as well as (super)exponential dependency on the treewidth, in practice these algorithms are often too slow to solve optimization problems. 
Though heuristic methods for finding tree decompositions of small width have been developed, most applications of tree decompositions are in speeding up exact algorithms. Little work has been done in using tree decompositions as a tool for high performance heuristic optimization algorithms.

To the best of our knowledge the only work in combinatorial optimization exploring this avenue is the tour merging algorithm for the Traveling Salesman Problem (TSP) by Cook and Seymour \cite{cook2003}, using the related concept of branch decomposition. In their paper they describe an algorithm that first generates a pool of high quality solutions to the TSP using a local search heuristic with different starting points. In the merging phase, the graph union of these solutions is then taken to produce a sparse subgraph of the original graph. This makes the computation of a low width branch decomposition feasible, which they then use to quickly find the optimal solution to the TSP instance induced by this sparse subgraph. Experimental results showed a fair improvement over the best solution found, in a small amount of additional time. 

In this paper we report experimental results applying the same paradigm described in \cite{cook2003} on the Steiner Tree Problem in Graphs (STP). A set of locally optimal solutions is generated to create a sparse subgraph, and subsequently tree decomposition is used to quickly solve the restricted instance to optimality. The main difference with the technique by Cook and Seymour is that we allow the algorithm to discard some of the generated solutions, if it helps finding a tree decomposition of sufficiently small width on the graph union of the remaining solutions. Though this hurts solution quality in some cases, the improvement in running time warrants this trade off. 

 For generating solutions we use a multistart heuristic by Ribeiro et al. \cite{ribeiro2002} available under the name \emph{Bossa}. The instances induced by the generated  solutions are solved using dynamic programming (DP), for which we use a fairly recent tree decomposition based implementation by Fafianie et al. \cite{fafianie2013}. We compare the performance of our algorithm to the \emph{path relinking} solution merging strategy proposed in \cite{ribeiro2002} which is part of the Bossa implementation. 

Experimental results show that our method is very promising.
Test runs on sparse benchmark sets showed up to an average 6-fold improvement of the optimality gap provided by the best generated solution, within only one or two percent of the  running time of the solution generating phase. On the other hand for dense graphs it often wasn't possible to find a combination of local solutions within our predefined treewidth limit. By using a fast greedy heuristic for finding tree decompositions however, it takes little time to identify this, and therefore the overhead of running the merging algorithm is negligible in such situations.

It should be noted that Bossa is no longer competitive in terms of performance. As pointed out by an anonymous reviewer, a heuristic by Polzin and Daneshmand \cite[Chapter~4]{polzin2003} was shown to give similar or better solutions compared to Bossa in a fraction of the time on established benchmarks. 

However, a very recent advancement by Pajor et al. \cite{pajor2014} indicates that our results are still highly relevant. They present an improved multistart heuristic  and experimental results indicate that this implementation outperforms the heuristic by Polzin and Daneshmand again. The proposed heuristic has a strong similarity to Bossa. Though some structural changes yield better quality solution pools in the same number of iterations, most of the performance gain is actually achieved by faster implementations of the same local search techniques. Since the improved implementation could be directly plugged in as a solution generator for our method, we expect the positive results to carry over when replacing the multistart heuristic with the improved version of Pajor et al., though further experiments are need to confirm this. 

The rest of this article is organized as follows. In Section \ref{sec:preliminaries} basic notation is introduced, we give a formal definition of treewidth and we discuss a greedy algorithm for treewidth that plays an important part in the performance of our algorithm. In Section \ref{sec: algorithm} we describe the heuristic for selecting solutions for merging and briefly discuss the algorithms used for generating solutions and solving the instance induced by those solutions. In Section \ref{sec: results} we report experimental results on a variety of benchmark instances.

\section{Preliminaries}
\label{sec:preliminaries}
We denote an undirected graph $G$ with vertex set $V(G)$ and edge set $E(G)$, or $V$ and $E$ when no confusion is possible. Together with a weight function $w : E \to \mathbb{R}$ we have a weighted graph. In this paper we assume graphs to be simple: no loops or parallel edges are allowed. A graph union $G \cup G'$ is equal to the graph found by taking the union of both the vertex and the edge sets of the operands. The neighbours of $v$ in $G$ are denoted $N_G(v) = \{u \in V(G) | (u,v) \in E(G)\}$.

The subject of this paper is the classical Steiner Tree Problem. This famous NP-Complete graph optimization problem should need no introduction to the reader but we include a formal definition for completeness: 
\begin{problem}[\textsc{Steiner Tree Problem}]\\
Given a connected weighted graph $G=(V,E,w)$ with non-negative weights and a set of terminal vertices $Q \subseteq V$, find a minimum weight subgraph $T$ of $G$ such that all terminal vertices are pairwise connected. 
\end{problem}

\subsubsection{Treewidth}
The concept of treewidth is a graph invariant that indicates how \emph{tree-like} a graph is. It is derived from the tree decomposition, a transformation that projects a general graph onto a tree. The formal definition is as follows:
\begin{definition}[Tree decomposition] A tree decomposition of a graph $G = (V,E)$ is a tree $\bar{T}=(I,F)$, where each node $i \in I$ is labelled with a vertex set $X_i\subset V$, called a bag, satisfying the following conditions: 
\begin{enumerate} 
\item 
$\bigcup_{i\in I} X_i = V$
\item 
for all $(v,w) \in E$ there is an $ X_i$ such that $ \{v,w\} \subset X_i$
\item
if $v \in X_i$ and $v \in X_j$ then $v \in X_k$ for all $k$ on the path between $i$ and $j$ in the tree $\bar{T}$
\end{enumerate} 
\end{definition}

 The width $w(\bar{T})$ of a tree decomposition is equal to the size of the largest bag minus one. The treewidth of a graph $\tw(G)$ is the smallest width over all possible tree decompositions of $G$. As finding the treewidth of a graph is NP-complete no polynomial time exact algorithms exists unless $P=NP$ \cite{bodlaender2006exact}.

 Heuristic approaches come in many shapes, including local search techniques and heuristics derived from exact algorithms, see Bodlaender and Koster \cite{bodlaender2008treewidth}. We will use a simple but very effective greedy heuristic described in \cite{bodlaender2008treewidth}.
 
 \begin{algorithm}
\caption{GreedyDegree(Graph $G$)}
\label{algo: greedydegree}
\begin{algorithmic}
\While{$|V(G)| \geq 0$}
\State $v \gets$ a minimum degree vertex in $G$
\State $\pi_i \gets v$
\State add edges to $G$ such that $N_G(v)$ is a clique
\State remove $v$ from $G$
\EndWhile
\State \Return $\pi = (\pi_1,...,\pi_n)$
\end{algorithmic}
\end{algorithm}

Algorithm \ref{algo: greedydegree}, {\sc GreedyDegree}, constructs an \emph{elimination order}, which is a permutation of vertices of the graph. It does so by iteratively choosing the minimum degree vertex, adding edges between all its neighbours, and then removing it from the graph. These last two steps are called \emph{vertex elimination}. 

Any elimination order can be used to construct a (unique) tree decomposition in linear time (see \cite[pg. 5]{bodlaender2008treewidth}). For convenience we will directly treat the output of Algorithm \ref{algo: greedydegree} as a tree decomposition. The width of the tree decomposition  produced by Algorithm \ref{algo: greedydegree} is equal to the highest degree of any vertex at the moment it is eliminated from the graph \cite{bodlaender2008treewidth}.  

In this paper we often abuse language by referring to the width found by \textsc{GreedyDegree}$(G)$ as the treewidth of $G$, especially when $G$ is a graph induced by a set of solutions. Of course this is just an upper bound, but since we never solve for exact treewidth in our algorithm,  it is not necessary to make the distinction when from context it is clear that we mean the width of the tree decomposition found.

\section{The algorithm}
\label{sec: algorithm} 
The basic outline of our approach consists of three steps. Let an instance of the STP be denoted by \textsc{STP}$(G,Q)$.

\begin{enumerate}
\item Generate as set $\mathcal{S}$ of locally optimal solutions for STP$(G, Q)$. 
\item Pick a subset $\mathcal{U} \subseteq \mathcal{S}$ such that $\bar{T}$ = \textsc{GreedyDegree}$(G_\mathcal{U})$ has width$(\bar{T})\leq m$, where $G_\mathcal{U} = \bigcup_{T\in\mathcal{U}} T$.
\item Solve STP$(G_\mathcal{U}, Q)$ using DP guided by the decomposition $\bar{T}$ found in 2.
\end{enumerate}

The DP implementation we used for the last step, more on that in Section \ref{subsec: dp}, has running time linear in $|V|$, but exponential in the treewidth. The multistart heuristic used to generate locally optimal solution is an implementation of a hybrid greedy randomized adaptive search procedure (GRASP) for the STP  (see section \ref{subsec: grasp}). As the first and last steps are basically black box routines with respect to the solution merging heuristic, we will first explain how we construct a suitable subset of solutions.

\subsection{Selecting Solutions}

In the implementation of tour merging for the TSP in \cite{cook2003} a fixed number of solutions is generated, and quite some time is spent on finding a good branch decomposition. If the algorithm can not find a decomposition of sufficiently small width, the merging heuristic is deemed intractable and returns no solution. 

Our method is a little different. We also generate a fixed number of heuristic solutions, and limit the width of the tree decomposition deemed acceptable to proceed with the DP step. However we allow more flexibility by  accepting a subset of solutions such that \textsc{GreedyDegree} finds a decomposition of width at most $m$ on their graph union. 

An initial approach to finding a good subset of solutions is motivated by the idea that if we cannot use all solutions, we give priority to those with the highest quality. Let $\mathcal{S}$ be the set of solutions generated in step 1 and $f(T) : S \to \mathbb{R}$ their weights. Initially we sort the solutions in ascending order of $f(T)$ and apply Algorithm \ref{algo: greedysteinerunion}. This keeps iteratively adding solutions to the graph union as long as the limit $m$ is not violated by the decomposition found by \textsc{GreedyDegree}. In a sense the algorithm finds a \emph{maximal} subset of solutions, that is, no solution can be added without breaching the width limit.

\begin{algorithm}
\footnotesize
\caption{\textsc{GreedySteinerUnion}$(\mathcal{S}, m)$}
\label{algo: greedysteinerunion}
\begin{algorithmic} 
\Require\\
	 $\mathcal{S}$:  List of Solutions, ordered\\
	$m$: Maximum treewidth  
\Ensure \\
$\mathcal{U}$: List of solutions, such that $tw\left(\bigcup_{T \in \mathcal{U}} T \right) \leq m$ 
\Procedure{GreedySteinerUnion}{$\mathcal{S}, m$}
\State $\mathcal{U} \gets \{\mathcal{S}(1)\}$
\For{$i = 2$ \textbf{to} $|\mathcal{S}|$}
\State $\mathcal{U'} \gets \mathcal{U} \cup \{S(i)\}$
\State $\ell \gets $ \Call{GreedyDegreeWidthMaxM}{$\bigcup_{T \in \mathcal{U'}} T, m$}
\If{$\ell \leq m$}
\State $\mathcal{U} \gets \mathcal{U'}$
\EndIf
\EndFor
\State \Return $\mathcal{U}$
\EndProcedure
\Procedure{GreedyDegreeWidthMaxM}{$G, m$} 
\State $\ell \gets 0$ 
\While {$\ell \leq m \wedge |V(G)| \geq 0$ }
	\State $v \gets$ a minimum degree vertex in $G$	
	\State $\ell \gets \max\{\textrm{degree}(v), \ell\}$
	\State add edges to $G$ such that $N_G(v)$ is a clique
	\State remove $v$ from $G$
\EndWhile
\State \Return $\ell$
\EndProcedure

\end{algorithmic} 
\end{algorithm}

This procedure usually gives reasonably good improvements in the DP step if the number of solutions rejected by Algorithm \ref{algo: greedysteinerunion} is low. 

However, if only  small sets of solutions stay within width limit $m$, and there are consequently many possible maximal solution sets, the chance of the greedy procedure finding a good set from the possible alternatives is small. Specifically, experiments showed that increasing the width limit $m$ may often result in a decrease in the eventual solution quality, a highly undesirable result. 

To improve the robustness of the solution picking step we introduce the randomized \emph{ranking} procedure described in Algorithm \ref{algo: rankingprocedure}. This procedure is akin to a simulation of step 2 and step 3 of the solution merging algorithm with a lower width limit $k$, where we shuffle the solutions instead of sorting them by $f(T)$. We use the value of the solution found in each iteration to adjust the rank of all solutions that were picked by Algorithm \ref{algo: greedysteinerunion} in that iteration. 

\begin{algorithm}
\footnotesize
\caption{\textsc{RankingProcedure}$(\mathcal{S}, f(T), k,r)$}
\label{algo: rankingprocedure}
\begin{algorithmic}
\Require\\
$\mathcal{S}$: List of Solutions\\
$f(T)$: Map giving the weigth of every Steiner Tree $T$ in $\mathcal{S}$\\
$k$: Maximum treewidth\\
$r$: Number of random ranking iterations
\Ensure\\
$f_A(T)$: Map assigning an adjusted value to every solution in $\mathcal{S}$
\Procedure{RankingProcedure}{$\mathcal{S}, f(T), k,r$}
\State $Z_T \gets \{f(T)\}, \forall T\in S$
\For{$r$ iterations}
\State Shuffle the order of $\mathcal{S}$ at random 
\State $\mathcal{U} \gets \textsc{GreedySteinerUnion}(\mathcal{S}, k)$
\State $\hat{z}\gets $ weight of Steiner Tree $T$ found by DP on the graph $G =\bigcup_{T \in \mathcal{U}} T$
\State add $\hat{z}$ to all sets $Z_T$ for which $T \in \mathcal{U}$
\EndFor
\State \Return $f_A(T) \gets \sum_{z' \in Z_T} \frac{z'}{|Z_T|}, \forall T \in \mathcal{S}$ 
\EndProcedure
\end{algorithmic}
\end{algorithm}

The adjusted values $f_A(T)$ can be interpreted as a metric for how promising the inclusion of a solution $T$ is in terms of the improvement found in step 3. These values are then used to sort the solutions before a final run of Algorithm \ref{algo: greedysteinerunion} with maximum width $m$. This yields a much more robust algorithm as in experiments we never observed an increase in $m$ resulting in a decrease in solution quality.

Experimental results indicate that the execution time of the DP grows roughly with $10^m$ where $m$ is the width of the tree decomposition. Therefore if we run Algorithm \ref{algo: rankingprocedure}, for example, with $k=m-2$ for 10 iterations, its execution time is still expected to be an order of magnitude smaller than directly running the DP once on a graph with decomposition of width $m$. A byproduct is that we can check more combinations of solutions for improvement. In fact, sometimes the best solution found during the execution of Algorithm \ref{algo: rankingprocedure} is better than the final solution found on the graph union with maximum width $m$, even after ranking according to the adjusted values. However, this does not happen too often and in general it pays off to execute a last iteration with the higher limit $m$. 

Taking it all together the steps for picking the set of solutions are: 
\begin{itemize}
\item find $f_A(T) = \textsc{RankingProcedure}(\mathcal{S}, f(T), k, r)$
\item sort the solutions $S$ ascending according to $f_A(T)$ 
\item find $\mathcal{U} = \textsc{GreedySteinerUnion}(\mathcal{S}, m)$
\end{itemize} 

The graph union of $\mathcal{U}$ and its tree decomposition are then used as input for the final DP run in step 3.

\subsection{Dynamic Programming} 
\label{subsec: dp}
A recent implementation of dynamic programming for the STP was introduced in \cite{fafianie2013}. It uses the {\sc GreedyDegree} algorithm to find a decent tree decomposition of the input graph, and then proceeds with a novel dynamic programming algorithm that reduces the search space in every stage by removing entries that cannot affect optimality. We will not reproduce the formal dynamic program here, for which we refer to the paper. 

However, the idea is that the DP is guided by a tree decomposition, such that the size of the state space is governed by the number of partitions of the vertex sets in each bag. In the paper multiple methods for reducing the size of the search space are proposed and implemented in the corresponding software. We use the default \emph{classic} DP however, as the relative speed ups are not large enough to make a significant contribution in our implementation.

\subsection{Greedy Randomized Adaptive Search Procedure}
\label{subsec: grasp}
A Hybrid for the STP was introduced in \cite{ribeiro2002} for which the code is publicly available under the name Bossa \cite{bossa}. 

Using a simple multistart approach, in which a construction heuristic is started from different nodes to produce a solutions that is then improved to a local optimum, does not work particularly well for the STP. For reasons that seem to be inherent to the problem most construction heuristics usually produce the same or a few different solutions even for widely different starting points. 

To still be able to improve on deterministic heuristics, the Hybrid GRASP algorithm in Bossa employs a variety of techniques to force the algorithm to explore different areas of the search space. These include multiple different construction heuristics, randomization in the local search procedure and weight perturbations. This makes the Hybrid GRASP particularly useful for our algorithm, as it can generate a set of good but disjoint solutions. For a full explanation of these techniques please see the paper. 

The Bossa code also includes a solution merging heuristic called Path Relinking, which can be used in combination with GRASP. We use it to compare the performance of our algorithm.

\section{Results} 
\label{sec: results}
The algorithm was implemented in JAVA integrating the existing JAVA code from \cite{fafianie2013} for the merging part and using system calls and text files to interface with the binary executable of Bossa, to generate the solutions pool. Though working with text files gave some overhead, this effect was insignificant as the time spent on read/write operations was usually small compared to the computation time. 

All experiments were run in a single thread on 16 core Intel Xeon E5-2650  v2 @ 2.6 GHz and 64GB of ram. At any time no more than 15 processes were running to make sure one core was free for background processes. The maximum heap space for the JAVA Virtual Machine was set to 1GB for all instances.

For all experiments, in the solution generation phase 16 solutions were generated with GRASP, where each run of the GRASP was set to 8 iterations and with a different random seed. We set the maximum treewidth for the final DP to $10$ and the maximum treewidth for the ranking procedure to $8$, with $20$ iterations of random shuffling. In all experiments where GRASP alone solved an instance to optimality, the instance was dropped from the test set.

\subsection{Benchmarks} 
\subsubsection{I640}

\begin{wraptable}{r}{0.6\textwidth}
\vspace{-30pt}
\caption{Results for I640} 
\label{tab: i640results}
\tiny \centering
\begin{tabular}{c l l | c  c  c | c  c  c | c }\\
 &&& \multicolumn{3}{c|}{Gap \% } & \multicolumn{3}{c|}{Time (s)} &  \\ 
Instance & $|Q|$ & $|E|$ & GRASP & SMH  & Impr. \% & GRASP & SMH & Rel. &  $\#$Trees\\
\hline \hline
201&50&960&0.28&0.28&0.0&9.4&2.1&0.2&13\\
204&50&960&0.44&0.00&100.0&8.8&0.4&0.1&5\\
205&50&960&0.02&0.02&0.0&8.0&4.0&0.5&11\\
211&50&4135&3.00&2.99&0.6&11.2&1.3&0.1&2\\
212&50&4135&3.23&2.95&8.7&11.2&3.0&0.3&2\\
213&50&4135&1.69&1.49&11.9&11.6&2.3&0.2&2\\
214&50&4135&2.22&2.22&0.0&9.2&5.1&0.6&2\\
215&50&4135&1.72&1.60&7.2&8.6&2.2&0.3&2\\
231&50&1280&0.09&0.01&92.3&5.1&6.7&1.3&6\\
232&50&1280&1.04&0.04&96.1&4.6&5.6&1.2&7\\
233&50&1280&0.62&0.00&100.0&5.0&6.2&1.2&6\\
234&50&1280&1.22&0.32&73.7&5.0&2.8&0.6&9\\
235&50&1280&0.74&0.39&46.8&4.5&4.7&1.1&3\\
241&50&40896&2.25&2.25&0.0&31.9&5.3&0.2&2\\
242&50&40896&1.79&1.79&0.0&31.6&4.5&0.1&2\\
243&50&40896&2.02&2.02&0.0&37.0&5.0&0.1&2\\
244&50&40896&1.62&1.62&0.0&30.8&5.1&0.2&2\\
245&50&40896&1.54&1.54&0.0&32.3&5.1&0.2&2\\
301&160&960&0.14&0.00&100.0&6.3&5.5&0.9&7\\
302&160&960&0.27&0.20&26.2&6.4&7.3&1.1&10\\
303&160&960&0.25&0.15&39.6&6.9&4.9&0.7&9\\
304&160&960&0.62&0.39&36.1&7.3&2.9&0.4&9\\
305&160&960&0.66&0.29&55.5&7.5&2.4&0.3&3\\
311&160&4135&1.49&1.49&0.0&12.6&1.7&0.1&1\\
312&160&4135&1.91&1.91&0.0&11.9&1.9&0.2&1\\
313&160&4135&1.49&1.49&0.0&12.0&1.9&0.2&1\\
314&160&4135&1.52&1.52&0.0&11.6&1.8&0.2&1\\
315&160&4135&1.63&1.63&0.0&12.0&1.3&0.1&1\\
331&160&1280&0.53&0.49&8.3&7.2&3.1&0.4&2\\
332&160&1280&0.84&0.84&0.0&8.1&2.0&0.2&2\\
333&160&1280&0.87&0.86&1.1&7.9&2.2&0.3&2\\
334&160&1280&1.00&1.00&0.0&6.7&1.5&0.2&1\\
335&160&1280&0.82&0.50&38.6&8.2&11.1&1.4&2\\
341&160&40896&0.70&0.70&0.0&68.3&3.0&0.0&1\\
342&160&40896&0.62&0.62&0.0&65.3&2.9&0.0&1\\
343&160&40896&0.53&0.53&0.0&60.2&2.7&0.0&1\\
344&160&40896&0.53&0.53&0.0&56.4&2.9&0.1&1\\
345&160&40896&0.60&0.60&0.0&58.8&1.9&0.0&1\\
\end{tabular}
\vspace{-20pt}
\end{wraptable}

An initial test was run on the last 50 instances of the classic I640 benchmark set available through the SteinLib \cite{steinlib} repository. All instances are randomly generated. This benchmark is a little outdated in that nowadays most instances can quickly be solved to optimality, but the clear distinction in parameters with which the instances were generated facilitates an easy analysis of the results. 

All instances in the benchmark set have 640 vertices, but differ in edge densities and number of terminals. For most instances the optimal value is known, in the other cases we used the best known upper bound as an approximation to find the optimality gap. This is only the case for instances I640-311$-$I640-315.

The results are in Table \ref{tab: i640results}. Next to the instance name the number of terminals and edges is shown. The optimality gaps of GRASP and our solution merging heuristic (SMH) are given as a percentage of the optimal value. The column Impr.\% gives the percentage improvement of the optimality gap by SMH compared to GRASP. The running time for SMH does not include GRASP. The column Rel. gives the time spent on SMH relative to the time spent on GRASP. The last column, \#Trees is the number of local solutions that were eventually accepted after the sorting procedure(see Algorithm \ref{algo: rankingprocedure}) in the solution union for the SMH.

The table clearly reveals the difference in performance between sparse and denser graphs. For instances with less than 1280 edges SMH usually gives a good improvement, even solving the instances to optimality in three instances, yet for none of the most dense instances an improvement was found. This is also reflected in the number of solutions that were used by the algorithm in the final run of the merging phase. There is a clear inverse relation between the density and the number of solutions the algorithm can merge while keeping treewidth within limits. As results e.g. instance 201 and 205 show, a high number of solutions merged does not guarantee improvement, although apparently it is a good indicator. Also the running time of the SMH relative to GRASP is usually lower when no improvement can be found.

As stated before most of the instances in the I640 are not particularly hard to solve with todays hardware. To get a better view of the power of the SMH algorithm we wanted to apply it to some bigger instances. The most notouriously hard test set in Steinlib is the PUC testset, of which most instances have no known optimal solution after more than 13 years in the field. No results are plotted but for completeness that SMH gave poor results on these instances: for all but the smallest instances we were not able to find any combination of solutions within width limit. We don't know if this is because our greedy tree decomposition works particularly bad for these graphs, or because high treewidth is an inherent property of the graph. In any case most instances from PUC are denser than the second highest density instances from I640, for which SMH was already hardly able to show improvement. 

Fortunately there are some other test sets in the Steinlib repository that are big enough to justify the use of our merging heuristic but not so dense as to make it run into trouble because of the treewidth limit. Results on these test sets are discussed in the rest of this section. 

To compare performance we also ran the path relinking algorithm (PR) from Bossa. The path relinking algorithm is itself a solution merging heuristic which comes in two flavours. On standard settings it first tries these different flavours and then picks the one that seems to perform best. For our experiments we forced it to use the random relink heuristic, as this turned out to perform best on all tested instances, and the initial run that determines the best settings takes a considerable amount of time. This makes for a more fair comparison. For more information see \cite{ribeiro2002}.

\subsubsection{ES1000(0)FST} 
The ES1000FST test set contains 15 instances  of randomly generated points on a grid, with L1 distances as edge weights. Each instance has 1000 terminal vertices and between 2500 and 2900 vertices in total. Due to preprocessing techniques applied to the graphs these instances only have between 3600 and 4500 edges, making them very sparse. 

The results for GRASP, GRASP+SMH and GRASP+PR are shown in Table \ref{tab: resultses1000fst}. Again, for SMH and PR the time does not include the initial GRASP iterations. Results are averaged over all instances. The number of instances for which the algorithms produced the best solution among all produced solutions for that instance is given by the row \#best.
\begin{table}
\centering
\begin{minipage}[b]{0.45\hsize}\centering
\caption{Results on ES1000FST} 
\label{tab: resultses1000fst}
\begin{tabular}{l |*{4}{l}}
&  & GRASP & SMH & PR \\\hline \hline
Opt Gap \%&&0.392&0.061&0.109\\
Time (s)&&402&6&493\\
\#Best&&0&14&1\\
\end{tabular}
\end{minipage}\hfill
\begin{minipage}[b]{0.45\hsize}\centering
\caption{Results on ES10000FST} 
\label{tab: resultses10kfst}
\begin{tabular}{l |*{3}{l}}
 && GRASP & SMH\\ \hline \hline
Opt Gap \%&&0.441&0.189\\
Total CPU Time(s)  &&194485&310\\
Wall Clock Time(s) &&12155& 310\\
\end{tabular}
\end{minipage}
\end{table}

Overall the SMH seems to perform better on these instances. Its also nice to note that on average 15.4 solutions were used in the final DP run of the merging phase. This is probably caused by the inherently low treewidth of the instances. However, the treewidth of these instances is not so low that direct use of DP would be feasible. As the treewidth found by \textsc{GreedyDegree} for the ES1000FST instances had a minimum of 14 and an average of 22, running the tree decomposition based DP on the original instances would take ages. 

We also tested on the single instance in the ES10000FST set. This graph is created in a similar way but with a factor 10 more terminals and vertices. Because this instance is so large that only running the 128 GRASP iterations needed for the SMH takes more than two cpu days, we did not compare it with path relinking in a sequential run. Instead grasp was run on 16 cores in parallel and the solution merging heuristic was run on a single core thereafter. Results are in Table \ref{tab: resultses10kfst}. Both the wallclock time (time untill all threads where finished) and the total computation time summed over all threads are shown. Though not the most spectacular improvement in optimality gap it shows the good scaling properties of SMH. The relative time spent on SMH compared to GRASP has about the same ratio as seen in the ES1000FST instances when we compare wallclock time, yet it is an order of magnitude smaller when we compare the total CPU time. We need to notice however that for this instance SMH was only able to use 3 solutions, as the treewidth of the entire solution pool combined was rather high at 22.  

\subsubsection{LIN}
The LIN test set from Steinlib has very similar properties as the ES1000FST set. These instances are generated from placing rectangles of different sizes in a plane, such that their corners become vertices and there edges graph edges. Though no preprocessing is done on these instances, it still makes for a very sparse graph, with no vertex having a degree more than 4. After dropping instances that were solved to optimality by GRASP, only the last 13 instances remained. The number of vertices of these instances are in the range 3700-39000.

\begin{table}

\centering
\begin{minipage}[b]{0.45\hsize}\centering
\caption{Results on LIN} 
\label{tab: resultslin}
\begin{tabular}{l |*{4}{l}}
  && GRASP & SMH & PR \\\hline \hline
Opt Gap \%&&0.33&0.09&0.13\\
Time(s)&&336&2&157\\
\#Best&&1&10&6\\
\end{tabular}
\end{minipage}\hfill
\begin{minipage}[b]{0.45\hsize}\centering
\caption{Results on ALUT/ALUE} 
\label{tab: resultsalu}
\begin{tabular}{l |*{4}{l}}
&  & GRASP & SMH & PR \\\hline \hline
Opt Gap \%&&0.27&0.07&0.11\\
Time(s)&&1386&7&1057\\
\#Best&&1&8&4\\
\end{tabular}
\end{minipage}

\end{table}

Results are in Table \ref{tab: resultslin}. Again the SMH performs very good compared to PR, in smaller amount of time. In all cases the merging phase could use all 16 solutions, often producing a union well below the treewidth limit. 

\subsubsection{ALUT/ALUE}
The last test sets we ran experiments on are the ALUT and ALUE sets  from Steinlib, which we combined because of their strong similarities. The structural properties of these instances are very much like the LIN test set. However, these come from very-large-scale-circuit (VLSI) applications. The results is a grid graph with rectangular holes in it. This graph again has a maximum degree of 4. After dropping instances which GRASP solved to optimality, 10  instances remained ranging in number of vertices between 3500 and 37000 and a number of terminals between 68 and 2344. Because of the fairly large size of some of these instances, we put a maximum on the running time for the combination of GRASP and merging heuristic of 3.5 hours. This gave a timeout for PR on the largest instance, so we took the best found solution up to that point. To compare, SMH only took 40 seconds to run for this instance, while GRASP took about 2 hours.

One of the nice properties of using the tree decomposition based approach is that for graphs with a regular structure such as with the last two sets we tested on, the size of the graph does not seem to matter much for the treewidth of the union of solutions, while the DP runs linear in the number of vertices. In the experiment run on the ALUT/ALUE sets, for all but two of the remaining instances the merging phase was able to use all 16 generated solutions. The two exceptions, where only 15 solutions were used, were the largest instance, and surprisingly, the smallest instance. This illustrates that observation quite well. 

\section{Conclusions and suggestions for further research}
Experimental results showed that a tree decomposition based exact algorithm can be employed as an efficient means to merge local heuristic solutions to the STP on sufficiently sparse graphs. As we have seen in results on the ALUT/ALUE test set, the sparse structure natural to VLSI derived graphs is exactly that at which our heuristic performs well. As VLSI is one of the major applications of the STP, this makes the heuristic practically relevant. 

As mentioned in the introduction the algorithm we used to generate solutions is no longer state of the art. In theory any algorithm capable of generating distinct locally optimal solutions could be employed with our algorithm. We plan to investigate the competitiveness of our solution merging heuristic when combined with a faster implementation such as \cite{pajor2014} for generating solutions in preparation of a journal version of this paper. 

That fixed parameter tractable algorithms can be used as a heuristic solution merging technique for the TSP had been shown in \cite{cook2003}, while we established results for the STP. It seems likely that there are more optimization problems where this technique can be used. A minimal requirement  seems to be that any feasible solution has low value for the chosen parameter. However, whether a low width decomposition can be found on a combination of local solutions depends on the instance, and in the case of the STP the density of the input graph seems a good indicator for that. It would be interesting to see if such a characterization is possible for other optimization problems that have low width solutions. 

As a final remark, in our algorithm we managed the treewidth of the solution union by discarding solutions. A simple extension would be to use an iterative scheme to reduce treewidth of the solution pool: first run the solution merging heuristic on (small) subsets of the generated solutions to generate a new solution pool with less solutions, and repeat until all solutions are within the treewidth limit. It seems likely this could further improve the performance.

\paragraph*{\textbf{Acknowledgments}}
I would like to thank N. Olver and L. Stougie for their feedback and an anonymous reviewer for helpful comments.

\end{document}